\documentclass[useAMS,usenatbib]{mn2e}

\usepackage{graphicx}
\usepackage{txfonts}
   \title{Evaporation of ices near massive stars: models based on laboratory TPD data}

\author[Viti et al.]{Serena Viti$^{1}$\thanks{sv@star.ucl.ac.uk}, 
          Mark P Collings$^{2}$,
	  John W Dever$^{2}$,
          Martin R S McCoustra$^{2}$,
          David A Williams$^{1}$ \\
$^1$ Department of Physics \& Astronomy, University College London, Gower St. WC1E 6BT U.K. \\
$^2$ School of Chemistry, University of Nottingham, University Park, Nottingham, NG7 2RD, UK. }
\begin{document}

   \date{Received ..; accepted ...}
\pagerange{\pageref{firstpage}--\pageref{lastpage}} \pubyear{2004}

\maketitle
\label{firstpage}

   \begin{abstract} Hot cores and their precursors contain an integrated
   record of the physics of the collapse process in the chemistry of
   the ices deposited during that collapse. In this paper, we present
   results from a new model of the chemistry near high mass stars in
   which the desorption of each species in the ice mixture is
   described as indicated by new experimental results obtained under
   conditions similar to those hot cores. Our models show that
   provided there is a monotonic increase in the temperature of the
   gas and dust surrounding the protostar, the changes in the chemical
   evolution of each species due to differential desorption are
   important. The species H$_2$S, SO,
   SO$_2$, OCS, H$_2$CS, CS, NS, CH$_3$OH, HCOOCH$_3$, CH$_2$CO,
   C$_2$H$_5$OH show a strong time dependence that may be a useful
signature of time evolution in the warm-up phase as the star moves on
to the Main Sequence.  This preliminary study demonstrates the consequences
   of incorporating reliable TPD data into chemical models.
\end{abstract}


\section{Introduction}

Conventionally, the first indirect manifestation of a young massive
star is provided by the presence of `hot cores'.  Hot cores are small,
dense, relatively warm, optically thick, and transient objects detected
in the vicinity of newly formed massive stars. It is well-known that
they exhibit a range of molecular species unlike those found in
quiescent molecular clouds: in comparison, hot cores are richer in
small saturated molecules and in large organic species (e.g. Walmsley
and Schilke 1993). The characteristic hot core chemistry is believed
to arise from the evaporation of the icy mantles that accumulated on
dust grains during the collapse that led to the formation of the
massive star; solid-state processing in the ice and gas-phase
chemistry after evaporation also occur (cf. Millar 1993; Nomura and
Millar 2004).  Moreover, recent observations do indicate the existence
of possible precursors of hot cores (e.g Molinari et al. 2002; Beuther
et al. 2002). Thus, hot cores and their precursors may be regarded as
containing an integrated record of the physics of the collapse process in the
chemistry of the ices deposited during that collapse.It is the task of
astrochemistry to unravel this record from the observations.

Viti \& Williams (1999) noted that the duration in which the grains
are warmed from very low ($\sim$10 K) temperatures to the temperature
at which H$_2$O ice desorbs and above is determined by the time taken
for a pre-stellar core to evolve towards the Main Sequence. This time
is believed from observational and theoretical studies to be fairly
brief, $\sim$10$^4$-10$^5$ y (e.g. Bernasconi \& Maeder 1996,
hereafter BM96), and in some contexts this `turn-on' of the star may
be regarded as instantaneous. However, in the context of hot cores,
which have a duration similar to that of the turn-on time of the hot
star, the relatively slow warming of the grains must be taken into
account. Evidently, highly volatile species will be desorbed from the
ices at early stages, while the more refractory species will only be
desorbed at later stages. The chemistry of the hot cores must reflect
this temperature-driven evolution, and the evolutionary chemistry
should provide, in principle, a direct measure of the stellar turn-on
time. Thus, hot cores contain not only a record of the collapse
process but also of the ignition history of the star. The
de-convolution of hot core observational data should constrain both
physical processes.

The Viti \& Williams study, in the absence of reliable data, made a
gross simplification in the description of desorption from mixed ices;
their paper emphasised that appropriate temperature programmed
desorption (TPD) experiments on mixed ices under near-interstellar
conditions were essential. Such definitive measurements have now been
made at the University of Nottingham Surface Astrochemistry
Experiment. Fraser et al. (2001) reported a TPD study of pure H$_2$O
ice, and Collings et al. (2003a) examined the TPD of H$_2$O/CO ice
mixtures. Both experiments provided data that is in striking contrast
to the assumptions made in models of hot cores. In particular,
Collings et al. showed that CO desorbed from an H$_2$O/CO ice mixture
occurs in four temperature bands, rather than the assumed CO
desorption at temperatures above about 25 K.  They also found that the
proportion of the CO desorbing at each phase depended on the total CO
concentration and on the preparation of the sample. These laboratory
results suggest that in hot cores the CO being injected into the gas
phase from the ices should occur in four phases, rather than a single
pulse.

The chemistry of interstellar ices is of course much more complex than
a simple H$_2$O/CO ice mixture, although H$_2$O is always the dominant
partner (Whittet 2003). Therefore, similar TPD experiments on a
variety of ices have been carried out at Nottingham (Collings et
al. 2004). In addition, the properties of some other species with
respect to desorption have been considered. It has been found possible
to categorize all species relevant to hot cores as H$_2$O-like
(i.e. behaving as determined by Fraser et al. 2001); CO-like (as
described by Collings et al. 2003a), or placed in several other 
intermediate categories. Thus, it is now possible to attempt to model
the chemistry near high mass stars in which the desorption of each
species in the ice mixture is described as indicated by experimental
results obtained under conditions similar to those hot cores. This
paper reports the results of such a study.

We model the chemistry of gas and the deposition of the ices during
the collapse, and subsequently the desorption and chemistry in the gas
phase after the star begins to warm the core. Our purpose is to
demonstrate the consequences of incorporating reliable TPD data into
chemical models. Specific sources will be modelled in later papers.

In Section 2 we describe the new TPD results. Section 3 gives
descriptions of the physical and chemical aspects of the model, and
the results are presented in Section 4. Section 5 gives a discussion
of the results and conclusions. For simplicity, throughout the paper,
we shall use the word `hot core' to include any pre-stellar core such
as those observed by Molinari et al. (2002 and references therein)
which are believed to be possible hot core precursors.


\begin{table*}
\caption{Assignment of each molecular species into a category of
desorption. The first column indicates the category, column 2 lists the species belonging to each category while columns 3-7 indicates the fraction of each species (remaining after the previous desorption) that desorbs at each desorption event. We have used italics for the experimentally
tested species.}
\begin{tabular}{|c|c|c|c|c|c|c|}
\hline
& Species&Solid &Mono &	Volcano&	Co-des&	Surface \\
& & at T$\sim$ 20K & at E$_b$ &  &   \\ 
\hline
CO-like &	{\it CO, N$_2$, O$_2$} &0.35&	0.7&	0.667&	1.0&	0 \\
	& CS, {\it NO, CH$_4$}&	0	&0.7&	0.667&	1.0&	0\\
Intermediate&	HCN, HNC,  {\it H$_2$S}, NS, HCS,{\it C$_2$H$_2$} &0&0.1&	0.5&1.0&0 \\
	& {\it OCS},{\it CH$_3$CN}, C$_3$H$_4$,{\it CO$_2$, SO$_2$}, HC$_3$N,{\it C$_2$H$_4$}& 0&0&0.5&1.0&0 \\
H$_2$O-like&	{\it H$_2$O, CH$_3$OH}, SO, CH$_2$CO, C$_2$H$_5$OH,{\it NH$_3$},NO$_2$, H$_2$CO,{\it HCOOCH}, H$_2$CS&0&0&0&1.0&0 \\
Reactive &HS$_2$, C$_2$H, OCN, O$_2$H, C$_2$H$_5$, H$_2$CN, HNO, C$_2$, C$_2$H$_3$&0&0&0&1.0&0 \\
Refractory& Mg, S$_2$&	0&	0&	0&	0&	1.0 \\
\hline

\end{tabular}
\label{tb:exp}
\end{table*}
\section{Experiments}

In the accompanying paper, Collings et al. (2004) report results of
TPD experiments for a large number of molecules relevant to hot core
chemistry. Details of these experiments are not discussed
here. Molecular were assigned to one of three several categories of
desorption as follows: (1) CO-like; (2) H$_2$O-like; and (3)
intermediate. The work also enables us to make inferences about likely
behaviour of some species not so far studied, and in doing so we have
created two further categories: (4) reactive; and (5) refractory. For
each category we have used the information in Collings et al. (2004),
and previous studies of the CO/H$_2$O system (Collings et al. 2003a,b) to
estimate the fraction of a particular molecular species that is
desorbed in the various temperature bands. These desorption bands are
from (i) the pure species; (ii) a mono-molecular layer on H$_2$O ice;
(iii) the amorphous to crystalline H$_2$O ice conversion (the `volcano'
effect); (iv) co-desorption when the H$_2$O ice desorbs; and (v)
desorption from the bare grain surface. The fraction of the species
desorbing in each of the five desorption processes is shown in Table
1, assuming a relatively thick ice layer of 0.3 $\mu$m, as used in the TPD
experiments. The desorption processes are tabulated in sequence, and
the quoted value expresses the fraction of the species remaining in
the ice at that time that is desorbed in the process.  In the layered
model of ice mantles, the species CO, N$_2$ and O$_2$, are thought to be
found predominantly in an outer layer non-hydrogenated (apolar) ice,
separate from the water dominated inner layer of hydrogenated ice
(polar). As the mantle warms from 10 K, the apolar layer is able to
diffuse into the porous structure of the water ice. We are yet to
accurately measure the relative concentration at which the water will
become saturated, but we estimate that this occurs at roughly
0.25. Since the total concentration of CO, N$_2$ and O$_2$ is our model is
0.35--0.4 relative to water, desorption of these species from the
non-hydrogenated outer layer will be relevant, and we list 35\% of each
as desorbed from the solid phase. Methane should only be found within
the hydrogenated inner layer, and we assume that this may also be true
of NO and CS. Based on the CO results of Collings et al. (2003b),
about 70\% of molecules of CO-like species in the hydrogenated layer
will undergo mono-molecular desorption in hot core conditions, and the
remainder will become trapped. The TPD results (Collings et al. 2004)
suggest that trapped molecules are released in volcano and
co-desorptions of equal sizes for the species initially present in the
water ice layer, and with a volcano desorption twice the size of the
co-desorption for the species that have diffused into the water
ice. The water-like species were found to be released entirely in the
co-desorption process. To the experimentally tested species we have
added polar organic molecules and other hydrophilic molecules. To the
intermediate category we have added untested non-polar organic species
and other small molecules that we anticipate will interact with have
moderate interaction with water ice. All of the intermediate species
are assumed to be found in the hydrogenated ice layer. The TPD results
(Collings et al. 2004), suggest that desorption will be roughly evenly
split between the volcano and co-desorption events. However, for the
smaller species C$_2$H$_2$, and H$_2$S, a small fraction ($\sim$ 10\%) is able to
escape and undergo mono-molecular desorption. We anticipate that HCN,
HNC, NS, and HCS will also behave in this manner.  The nine species we
have categorised as reactive are all ions or free radicals in the gas
phase. It is not clear how they will behave when adsorbed in the grain
mantle at 10 K, however as the ice warms they would be expected to
adopt a form quite different from their gas phase structure due to
strong interactions with the water ice and other molecules. As with
the water-like species, we expect that they will only escape to the
gas phase in a co-desorption process. The refractory species are those
such as atomic magnesium and S$_2$ which are likely to be stable in the
mantle but are strongly adsorbed and non-volatile. We intuitively
expect that they will remain in the mantle as other species, including
water, desorb around them. Unless some explosive desorption mechanism
operated, we anticipate that they will 'sink' through the desorbing
mantle and become bound to the dust grain surface. Desorption from the
dust grain surface is therefore the only relevant process for this category.

\section{The chemical and physical model}

The model employed is similar to that used in VW99 and Viti et
al. (2001). For this study, we performed time-dependent single-point calculations for A$_V$ of $\sim$ 600 mags as derived from hot
core observations. For simplicity
we have not used a multi-point code, as we did in
VW99. 
Our model is a two phase calculation: the first phase starts
from a fairly diffuse ($\sim$ 300 cm$^{-3}$) medium in atomic form
(apart from a fraction of hydrogen in H$_2$), and undergoes a
free-fall collapse until densities typical of hot cores are reached
($\sim$ 10$^7$ cm$^{-3}$). During this time, atoms and molecules from the gas freeze onto the
grains and they hydrogenate where possible, as in VW99. 
Note that the advantage of this approach is that the ice composition is $not$ assumed but it is derived by a time
dependent computation of the chemical evolution of the gas/dust interaction process. Also the abundances that we derive for the high density core are directly computed from the initial cosmic abundances.
The main differences here with the VW99 models are
i) the initial abundances which we corrected to match the latest
findings by Sofia et al (2001) and are 1.0, 0.075,
4.45$\times$10$^{-4}$, 1.79$\times$10$^{-4}$, 8.52$\times$10$^{-5}$,
1.43$\times$10$^{-6}$, 5.12$\times$10$^{-6}$, respectively, for H, He,
O, C, N, S, Mg; ii) the use of the UMIST rate 99 database (www.rate99.co.uk) rather
than UMIST96; iii) the low initial density (cf. 10$^4$ cm$^{-3}$ in VW99). The
unusually high initial density in VW99 before collapse was employed
for ease of computation.  Our present models start from a more
realistic low density diffuse medium and this has consequences for the final
percentage of gas in the mantles. In VW99, effectively 100\% of the
gas species was frozen onto the grains once the final density was
reached. However, in Viti et al 2001, we argued that it is possible that freeze out is never total;
here we assume that 99\% of the gas is frozen out, leaving 1\% in the gas (a more realistic
description since some gaseous CO is always observed, even in regions
where no millimetre continuum is detected, see e.g. Molinari et
al. 2000). This difference affects the very start of Phase 2 of our
computations. These differences do not affect at all
our qualitative conclusions.  We also performed a model where the initial
abundances and rate file used matched the VW99 for an initial
comparison (see Section 4.1). 

\par Phase 2 follows the chemical evolution
of the remnant core. We simulate the effect of the presence of an
infrared source in the centre of the core or in its vicinity by
subjecting the core to an increase in the gas and dust temperature.
The temperature reaches its maximum ($\sim$ 300K) at different times
depending on mass of the new born star (see VW99).

Phase 2 is rather different from that in the VW99 models because of
the different treatment for the increasing temperature and the
evaporation from the grains.In VW99 we adopted a linear dependence of
the temperature with time. For this study we have opted for a more
realistic approach where we derived the temperature of the gas as a
function of the luminosity (and therefore the age) of the
protostar. We used the observational luminosity function of Molinari
et al. (2000); in their Figure 6 the total emitted luminosity is
plotted against the core mass for mass accretion rates of 10$^{-5}$
and 10$^{-4}$ solar masses yr$^{-1}$. Using this plot we correlated
the luminosity, and hence the effective temperature, with the age of
the accreting protostar and found that a power law fitted the
data. Our major assumption was then that the temperature of the gas
and dust surrounding the accreting protostar increases according to
the same power law as the stellar temperature, and we fitted this so
that the maximum temperature reached by the gas at the contraction
time, t$_c$, i.e the time after which hydrogen starts burning and the
star reaches the ZAMS; Table~\ref{tb:tc_temps} lists the contraction times and
the volcano and co-desorption temperatures as a function of the mass of the stars. 
The contraction times are taken from 
BM96. Note, that the precise contraction time and the exact
behaviour of the temperature increase are mostly irrelevant to the purpose of
this study: the only necessary assumptions are that i) the contraction
time of a massive protostar is comparable to chemical timescales - all
the evolutionary studies of the type performed by BM96 have shown that
even the most massive stars ($\sim$ 120 M$_{\odot}$) contract in no
less than 10,000 years and ii) the temperature increases monotonically
and not following a step function. It is true that the volcano and the co-desorption temperatures 
decrease as a function of the slowing down of the heating rate so that the lower the 
mass of the star, the lower the volcano and the co-desorption temperatures (see Table~\ref{tb:tc_temps}) and Collings et al. 2004) however this decrease is very small and insignificant with respect of the evaporation times.

Our new evaporation treatment is pseudo-time dependent in that we
allow evaporation of a fraction of mantle species $X$ (in a single
step) when the temperature for a particular desorption event is
reached. The temperature and fractions for each species are listed in
Table 1. Note that the fractions are always expressed as fractions of
the current amount and not the original.
\begin{table}
\caption{Contraction times ($t_c$)  and
the volcano and co-desorption temperatures as a function of the mass of the stars.}
\begin{tabular}{|cccc|}
\hline
M  & $t_c$ & Volcano Desorption & co-desorption \\ 
M$_{\odot}$ & 10$^6$ years & Kelvin & Kelvin \\
\hline
60 & 0.0282 & 92.2  & 103.4 \\
25 & 0.0708 & 90.4 & 101.6 \\
15 & 0.117 & 89.5 & 100.8 \\
10 & 0.288 & 88.2 & 99.4 \\
5 & 1.15 & 86.3  & 97.5 \\
\hline
\end{tabular}
\label{tb:tc_temps}
\end{table}

\section{Results}
The first task of this theoretical study is to determine whether the
inclusion of the five distinct desorption events occurring during TPD of dirty
ice affect significantly the abundances predicted by
previous models, in particular the VW99 models. To isolate this effect
we ran several models using the same initial conditions of VW99 i.e
the same initial abundances (see Table 2 from VW99) and chemical database. 
We found that substantial differences are seen between these models
and VW99. These differences are solely due to the new evaporation
treatment.
\par
We then constructed a new grid of models using the revised initial
elemental abundances listed in Section 3, and the latest database
UMIST99.  We ran several models covering stars with masses from 5 to
60 M$_{\odot}$. We have taken the contraction times from BM96 as the
epochs at which the temperature of the core reaches 300K. The
desorption temperatures were calculated by running simulations of
water ice desorption (Collings et al. 2003b) using the power law
temperature profiles fitted to the stars of each mass.  Figures 1-3
show the fractional abundances obtained from our computations for
selected species for respectively a M = 25 M$_{\odot}$, M = 5
M$_{\odot}$, and M = 15 M$_{\odot}$ star.
\begin{figure}
\vspace{8cm}
\includegraphics{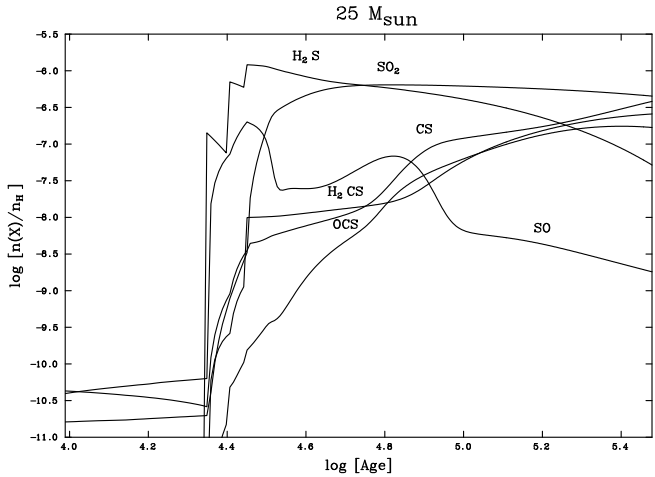}
\caption[]{Fractional abundances of selected sulfur-bearing species as a function of time for a 25 M$_{\odot}$ star, equivalent to a contraction time of 70,000 years.}
\label{fg:fig1}
\end{figure}
\begin{figure}
\vspace{8cm}
\includegraphics{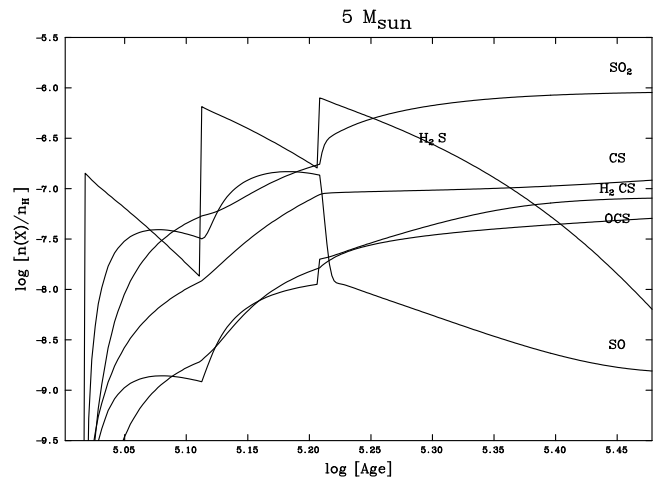}
\caption[]{As in Figure 1 but for a 5 M$_{\odot}$ star, equivalent to a contraction time of 1.15$\times$10$^6$ yrs.}
\label{fg:fig2}
\end{figure}
\begin{figure}
\vspace{8cm}
\includegraphics{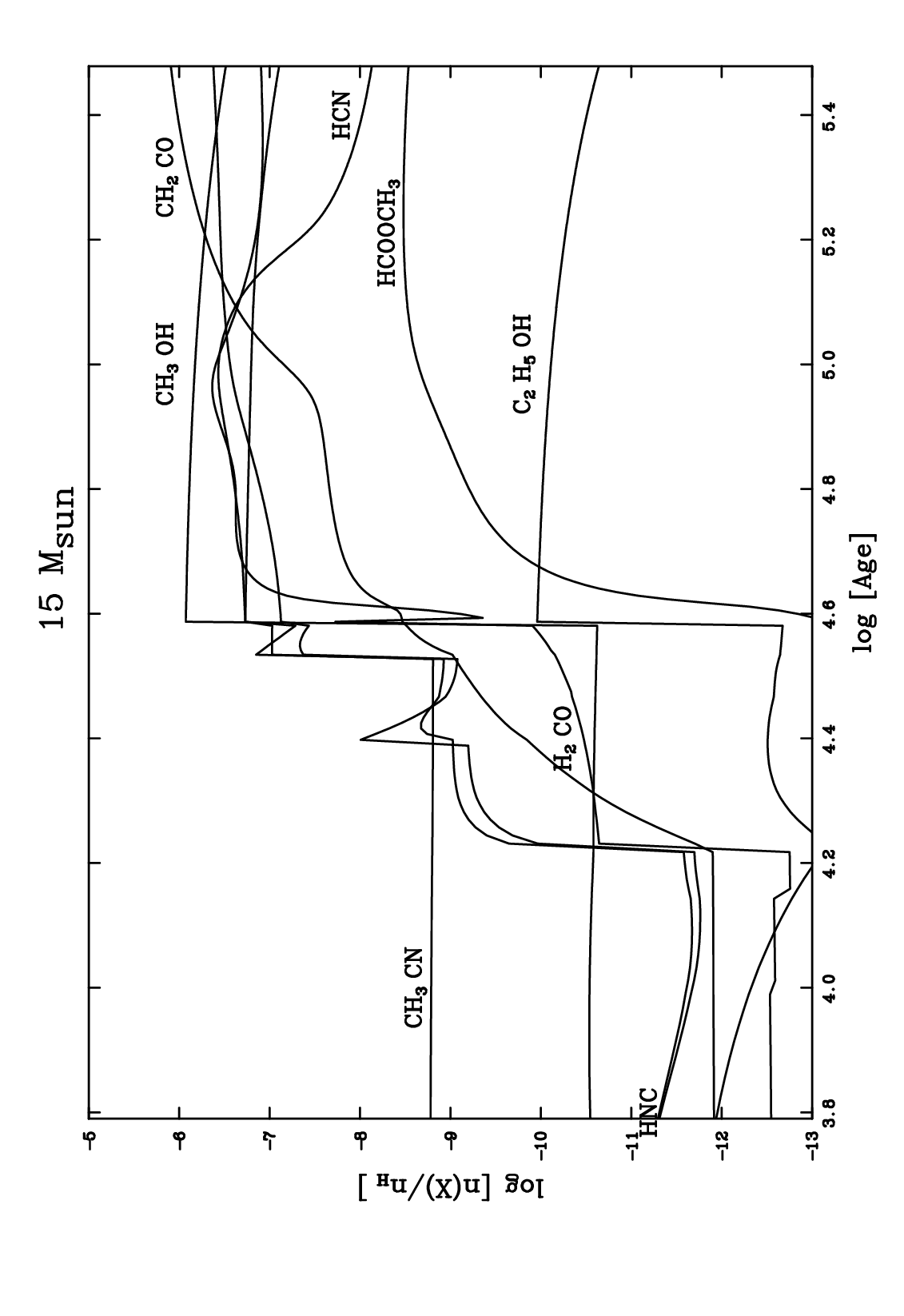}
\caption[]{Fractional abundances of selected large molecules as a function of time for a 15 M$_{\odot}$ star, equivalent to a contraction time of 70,000years.}
\label{fg:fig3}
\end{figure}

The species that are most affected by the inclusion of the new
evaporation treatment are sulfur-bearing species and the larger species.
The sensitivity of sulfur-bearing species to
physical and chemical variations during the lifetime of a hot cores is
not surprising as previous studies have shown that sulfur-bearing
species are strongly time-dependent (e.g. Hatchell et al. 1998; Viti et
al. 2001, Hatchell \& Viti 2002). While larger molecules, such as CH$_3$OH and CH$_2$CO,
maintain the same trends from a 5 to a 25 M$_{\odot}$ star, the behaviour of 
sulfur-bearing species depend on the heating rate and hence on the mass of the star. 
In particular, we find that the
evaporation of H$_2$S at different evolutionary stages is one of the main key factor to
the changes occurring in the sulfur bearing chemistry: for a 
25 M$_{\odot}$ star 
 the abundance of
H$_2$S itself varies by about one order of magnitude with time;
as the mass of the star decreases 
this difference becomes larger, up to
two orders of magnitude.
 H$_2$S has two main peaks
in abundance due to desorption from the volcano and co-desorption
events, but each peak is followed by a quick and sharp decline due to
dissociation. This fast and changing behaviour has several
consequences on other sulfur bearing species; it is interesting to
look at these changes in the ratios of selected sulfur bearing
species e.g: for a 25 M$_{\odot}$ star, (i) H$_2$S/SO$_2$ is $>$ 1 up to 50,000 years, and $<$ 1 for
older cores.  (ii) H$_2$S/CS, H$_2$S/H$_2$CS, H$_2$S/OCS $>$ 1 up to 150,000 yrs and $<$ 1 for older cores;
(iii) SO/CS, SO/H$_2$CS, SO/OCS $\sim$ 1 up to 65,000 yrs and $<$ 1 after that. For a 5 M$_{\odot}$ star,
we find similar trends but much more accentuated; for example, H$_2$S, SO and SO$_2$ show similar abundances for longer periods during the earlier phases of star formation, while H$_2$S/SO$_2$ is much less than 1 
after 190,000 yrs. In general H$_2$S decreases much more drastically with time as we move toward
the lower mass regime; this is because, although the temperatures
at which the evaporations occur are quite similar between the 5 and the 25 M$_{\odot}$ cases,
the temperature increases occurs more slowly for a 5 M$_{\odot}$ star, allowing
more time for the gas H$_2$S to be destroyed, before another H$_2$S evaporation `peak' occurs. 
The main destruction
route for H$_2$S is its reaction with HCO$^+$; the latter becomes abundant once the temperature reaches
$\sim$ 21 K and the first CO evaporation peak occurs.   
From these trends, it is clear that the ratio of these sulfur-bearing
species can be used to investigate the evolution of the early phases
of high mass star formation, and that sulfur bearing species are therefore good `chemical
clocks'.
\par
Another set of species that can be used as evolutionary indicators are organic molecules. Figure 3
shows the time evolution for selected large species for a 15 M$_{\odot}$ star. 
From the figure, we note that 
large species, such as C$_2$H$_5$OH, HCOOCH$_3$ and CH$_3$OH, are
good indicators of old cores as these strongly bound species are
abundant in the gas only at late times.
\par

Although observationally less significant, we would also like to draw
attention to other species (not shown) namely: CO varies by about 1
order of magnitude during the increase of the temperature and
therefore during the formation of a high mass star but, CO being a
ubiquitous species, this difference, even if detectable, may not, by
itself, be indicative. Finally, we find that N$_2$H$^+$ should be
present in detectable quantities in young hot cores but disappears
probably within the first 15,000 years of its lifetime, when the temperature has reached $\sim$ 40K; this corresponds 
roughly to the mono evaporation of CO ($\sim$ 40 K) and of CH$_4$ ($\sim$ 45 K). In fact the main destruction routes for N$_2$H$^+$ are its reactions
with methane and CO.  
 On the contrary ammonia should
only be detectable in cores older than 30,000 years (for a 25 M$_{\odot}$)
Intermediate aged high mass star formation cores
should be in general nitrogen-deficient.

\section{Conclusions}
We have extended existing chemical models of high mass star formation cores by including
the latest experimental results on desorption from grains in the
evaporation of icy mantles formed in star-forming regions. We find
that distinct chemical events occur at specific grain temperatures.
This supports our assertion that in a comprehensive model it should be possible to use 
these chemical events as
a chemical clock tracing the ignition of the nearby star.

We have shown that, provided there is a monotonic increase in the
temperature of the gas and dust surrounding the protostar, the changes
in the chemical evolution of each species due to differential
desorption are important. Based on the recent TPD experimental results, 
the species that are most affected are:
H$_2$S, SO, SO$_2$, OCS, H$_2$CS, CS, NS, CH$_3$OH, HCOOCH$_3$,
CH$_2$CO, C$_2$H$_5$OH, and to a lesser extent CO, N$_2$H$^+$, NH$_3$
and HCN, HNC. Our results imply that if a large enough sample of high
mass precursors objects are observed in these molecules, we would
gather enough information to determine their age and therefore the
evolutionary stage of the high mass star formation core.

Finally, we note that
1) effects induced by the relatively slow warming of gas near 
stars are more pronounced near stars of moderate mass, since
the turn-on time for stars of intermediate mass is much longer than for
massive stars. Observations of cores similar to hot cores are now
being found for intermediate-to-low mass stars (e.g Cazaux et al. 2003); 
2) the evolutionary trends found in this study depend on the temperature
gradient. Hence, at the established `hot core' stage, similar trends should be found
as a function of distance from the central star, again due to a temperature 
gradient
(e.g Nomura and Millar 2004).

\section{acknowledgements}
SV acknowledges individual financial support from a PPARC Advanced Fellowship.
MPC acknowledges the
support of the Leverhulme Trust. DAW is grateful for the award of the Leverhulme Trust Emeritus Fellowship. JWD
acknowledges the individual financial support of a PPARC studentship. Mark Anderson and Rui Chen are thanked for helpful discussions.

\label{lastpage}
\end{document}